\documentclass[fleqn,twoside]{article}
\usepackage[headings]{espcrc2}
\usepackage{booktabs}

\makeatletter
\renewenvironment{thebibliography}[1]
     {\section*{REFERENCES}%
      \@mkboth{\MakeUppercase\refname}{\MakeUppercase\refname}%
      \list{\@arabic\c@enumiv.}%
           {\settowidth\labelwidth{#1.}%
            \leftmargin\labelwidth
            \advance\leftmargin\labelsep
            \@openbib@code
            \itemsep\z@ \parsep\z@
            \usecounter{enumiv}%
            \def\makelabel##1{\rlap{##1}\hss}%
            \let\p@enumiv\@empty
            \renewcommand\theenumiv{\@arabic\c@enumiv}}%
      \sloppy
      \clubpenalty4000
      \@clubpenalty \clubpenalty
      \widowpenalty4000%
      \sfcode`\.\@m}
     {\def\@noitemerr
       {\@latex@warning{Empty `thebibliography' environment}}%
      \endlist}
\makeatother

\readRCS
$Id: espcrc2.tex,v 1.2 2004/02/24 11:22:11 spepping Exp $
\usepackage{graphicx}
\usepackage[figuresright]{rotating}

\newcommand{ \be}{\begin{equation}}
\newcommand{ \ee}{\end{equation}}
\newcommand{ \bea}{\begin{eqnarray}}
\newcommand{ \eea}{\end{eqnarray}}
\newcommand{ \mysmall}[1]{\scriptscriptstyle #1} 

\newcommand{ \amu}{a_{\mu}}
\newcommand{ \mw}{M_{\mysmall{W}}}
\newcommand{ \mz}{M_{\mysmall{Z}}}
\newcommand{ \mh}{M_{\mysmall{H}}}

\newcommand{ \eq}[1]{Eq.~(\ref{eq:#1})}
\newcommand{ \gev}  {\mbox{ GeV}}
\newcommand{ \bm}   {\boldmath}
\newcommand{ \ubm}  {\unboldmath}

\newcommand{\AmS}{{\protect\the\textfont2
  A\kern-.1667em\lower.5ex\hbox{M}\kern-.125emS}}

\title{Electron, muon and tau magnetic moments: a theoretical update}

\author{M.~Passera\address{Dipartimento di Fisica, Universit\`a di Padova
and INFN, Via Marzolo 8, 35131 Padova, Italy}\thanks{Work supported
in part by the European Community's Marie Curie Research Training Networks
under contracts 
MRTN-CT-2004-503369 and
MRTN-CT-2006-035505. }}
       
\begin{document}

\begin{abstract}
Recent Standard Model predictions for the anomalous magnetic moments of the
electron, muon and $\tau$ lepton are reviewed and compared to the latest
experimental values.
\end{abstract}

\maketitle

\section{INTRODUCTION}
\label{sec:INTRODUCTION}

The gyromagnetic factor $g$ is defined by the relation between the
particle's spin $\vec{s}$ and its magnetic moment $\vec{\mu}$,
\be
\vec{\mu}=g \frac {e} {2m} \vec{s},
\ee
where $e$ and $m$ are the charge and mass of the particle.  In the Dirac
theory of a charged point-like spin-$1/2$ particle, $g=2$.  Quantum
Electrodynamics ({\small QED}) predicts deviations from Dirac's value, as
the charged particle can emit and reabsorb virtual photons.  These {\small
QED} effects slightly increase the $g$ value. It is conventional to express
the difference of $g$ from 2 in terms of the value of the so-called
anomalous magnetic moment, a dimensionless quantity defined as $a = (g -
2)/2$.

The measurements of the anomalous magnetic moment of the electron and the
muon recently reached the fabulous relative precision of 0.7 parts per
billion (ppb)~\cite{Gabrielse_g_2006} and 0.5 parts per million
(ppm)~\cite{BNL,BNL04-6}, respectively.  The theoretical prediction of the
electron one, $a_{e}$, is rather insensitive to strong and weak
interactions, hence providing a stringent test of {\small QED} and leading
to the most precise determination of the fine-structure constant $\alpha$ to
date~\cite{Gabrielse_a_2006,MP06}. On the other hand, the muon anomalous
magnetic moment $a_{\mu}$ allows to test the entire Standard Model ({\small
SM}) and scrutinize viable alternatives to this theory, as each of its
sectors contributes in a significant way to the total prediction. Compared
with $a_e$, $a_{\mu}$ is also much better suited to unveil or constrain
``New Physics'' ({\small NP}) effects~\cite{CM01}. Indeed, for a lepton $l$,
their contribution to $a_l$ is generally expected to be proportional to
$m_l^2/\Lambda^2$, where $m_l$ is the mass of the lepton and $\Lambda$ is
the scale of {\small NP}, thus leading to an $(m_{\mu}/m_e)^2 \sim 4\times
10^4$ relative enhancement of the sensitivity of the muon versus the
electron anomalous magnetic moment. This more than compensates the much
higher accuracy with which the $g$ factor of the latter is known.  The
anomalous magnetic moment of the $\tau$ lepton, $a_{\tau}$, would suit even
better; however, its direct experimental measurement is prevented by
the relatively short lifetime of this lepton, at least at present.
The existing limits are based on the measurements of the
total and differential cross sections of the reactions $e^+e^- \to
e^+e^-\tau^+\tau^-$ and $e^+e^- \to Z \to \tau^+\tau^-\gamma$ at {\small
LEP} energies. The most stringent limit, $-0.052 < a_{\tau}^{\mysmall \rm
EXP} < 0.013$ at 95\% confidence level, was set by the {\small DELPHI}
collaboration~\cite{delphi}, and is still more than an order of magnitude
worse than that required to determine $a_{\tau}$.

The {\small SM} prediction $a_{l}^{\mysmall \rm SM}$ ($l\!=\!e$, $\mu$ or
$\tau$) is usually split into three parts: {\small QED}, electroweak
({\small EW}) and hadronic. Here we provide a summary of the present status
of these contributions, and a comparison with the current experimental
values.

\section{QED EFFECTS: PREAMBLE}
\label{sec:QED}

The {\small QED} part of the anomalous magnetic moment $a_l$ of a charged
lepton $l\!=\!e$, $\mu$ or $\tau$ arises from the subset of {\small SM}
diagrams containing only leptons and photons. For each of the three leptons
$l$, of mass $m_l$, this dimensionless quantity can be cast in the general
form~\cite{KM90}
\be
    a_l^{\mysmall \rm QED} \!\!=\! A_1 \!+ 
                   A_2 \!\!\left(\!\frac{m_l}{m_j} \!\right) \!+ 
                   A_2 \!\!\left(\!\frac{m_l}{m_k} \!\right) \!+ 
                   A_3 \!\!\left(\!\frac{m_l}{m_j},
                   \!\frac{m_l}{m_k}\!\right)\!,    
\label{eq:amuqedgeneral}
\ee
where $m_j$ and $m_k$ are the masses of the other two leptons. The term
$A_1$, arising from diagrams containing photons and leptons of only one
flavor, is mass and flavor independent.  In contrast, the terms $A_2$ and
$A_3$ are functions of the indicated mass ratios, and are generated by
graphs containing also leptons of flavors different from $l$. The
contribution of a lepton $j$ to $a_l^{\mysmall \rm QED}$ is suppressed by
$(m_l^2/m_j^2)$ if $m_j \!\gg\! m_l$, while it contains a logarithmic
enhancement factor $\ln(m_l/m_j)$ if $m_j \!\ll\! m_l$.  The muon
contribution to $a_e^{\mysmall \rm QED}$, for example, is thus power
suppressed by a factor $(m_e^2/m_{\mu}^2) \sim 10^{-5}$; nonetheless, this
effect is much larger than the tiny experimental uncertainty on
$a_e$~\cite{Gabrielse_g_2006} (see Sec.~\ref{sec:ELECTRON}).  On the
contrary, the {\small QED} parts of $a_{\mu,\tau}$ beyond one-loop are
dominated by the mass-dependent terms.

The functions $A_i$ ($i\!=\!1,2,3$) can be expanded as power series in
$\alpha/\pi$ and computed order-by-order
\be
    A_i \!=\! A_i^{(2)}\!\left(\frac{\alpha}{\pi} \right)
    + A_i^{(4)}\!\left(\frac{\alpha}{\pi} \right)^{\!2}
    + A_i^{(6)}\!\left(\frac{\alpha}{\pi} \right)^{\!3} +\cdots.
\ee
Only one diagram is involved in the evaluation of the lowest-order
(first-order in $\alpha$, second-order in the electric charge)
contribution---it provides the famous result by Schwinger $A_1^{(2)} \!=\!
1/2$~\cite{Sch48}. The mass-dependent coefficients $A_2$ and $A_3$ are of
higher order. All results discussed below were derived using the latest
{\small CODATA}~\cite{CODATA02} recommended mass ratios: $ m_e/m_{\mu} =
4.836 \, 331 \, 67 (13) \times 10^{-3}$, $ m_e/m_{\tau} = 2.875 \, 64 (47)
\times 10^{-4}$, $ m_{\mu}/m_e = 206.768 \, 2838 (54)$, $ m_{\mu}/m_{\tau} =
5.945 \, 92 (97) \times 10^{-2}$, $ m_{\tau}/m_e = 3477.48 (57)$, $
m_{\tau}/m_{\mu} = 16.8183 (27)$. The value for $m_{\tau}$ adopted by
{\small CODATA} in Ref.~\cite{CODATA02} ($m_{\tau}= 1776.99\, (29)$ MeV) is
based on the 2002 {\small PDG} result~\cite{PDG02}. This {\small PDG} result
remains unchanged to date~\cite{PDG04,PDG06}.

\section{ELECTRON}
\label{sec:ELECTRON}

\subsection{QED Contributions}

Seven diagrams contribute to the fourth-order coefficient $A_1^{(4)}$, one
to $A_2^{(4)}(m_e/m_{\mu})$ and one to $A_2^{(4)}(m_e/m_{\tau})$. As there
are no two-loop diagrams contributing to $a_e^{\mysmall \rm QED}$ that
contain both virtual muons and taus, $A_3^{(4)}(m_e/m_{\mu},m_e/m_{\tau})
\!=\! 0$. The mass-independent coefficient has been known for almost fifty
years~\cite{So57-58-Pe57-58}:
\be
    A_1^{(4)} = -0.328 \, 478 \, 965 \, 579 \, 193 \, 78 \ldots.
\label{eq:A14}
\ee

The coefficient of the two-loop mass-dependent contribution to
$a_l^{\mysmall \rm QED}$, $A_2^{(4)}(1/x)$, with $x\!=\!m_j/m_l$, is
generated by the diagram with a vacuum polarization subgraph containing the
virtual lepton $j$.
This coefficient was first computed in the late 1950s for the muon $g$$-$$2$
with $x \!=\! m_e/m_{\mu} \!\ll\! 1$, neglecting terms of
$O(x)$~\cite{SWP57}. The exact expression for $0\!<\!x\!<\!1$ was reported
by Elend in 1966~\cite{El66}. However, for many years its numerical
evaluation was considered tricky because of large cancellations and
difficulties in the estimate of the accuracy of the results, so that common
practice was to use series expansions
instead~\cite{Samuel91,Samuel93,CS99}. Taking advantage of the properties of
the dilogarithm ${\rm Li}_2(z)=-\!\int_0^z (dt/t) \ln(1-t)$~\cite{Lewin},
the exact result was cast in~\cite{MP04} in a very simple and compact
analytic form, valid, contrary to the one in~\cite{El66}, also for $x \!\geq
\! 1$ (the case relevant to $a_e^{\mysmall \rm QED}$ and part of
$a_{\mu}^{\mysmall \rm QED}$). Its numerical evaluation with the mass ratios
given in Sec.~\ref{sec:QED} yields~\cite{MP06}
\bea
     A_2^{(4)} (m_e/m_{\mu})  
     & = & 5.197 \, 386 \, 70 \, (28) \times 10^{-7} 
\label{eq:EA24m}
\\
     A_2^{(4)} (m_e/m_{\tau})  
     & = &  1.837 \, 62 \, (60) \times 10^{-9}, 
\label{eq:EA24t}
\eea
where the errors are only due to the uncertainties of the mass ratios. The
results of Eqs.~(\ref{eq:EA24m}) and (\ref{eq:EA24t}) are equal to those
obtained with a series expansion in powers of the mass ratio $y$ and $\ln
y$, with $y\!  \ll \!1$~\cite{CODATA02}.

Adding up Eqs.~(\ref{eq:A14}), (\ref{eq:EA24m}) and (\ref{eq:EA24t}) one
gets the two-loop {\small QED} coefficient~\cite{MP06}
\bea
    C^{(4)}_e &=& A_1^{(4)} + A_2^{(4)}(m_e/m_{\mu}) + 
               A_2^{(4)}(m_e/m_{\tau}) 
     \nonumber \\     &=& 
     - 0.328 \, 478 \, 444 \, 002 \, 90 \, (60).
\label{eq:EC2}
\eea 
The mass-dependent part of $C_e^{(4)}$ is small but not negligible, giving a
relative contribution to the theoretical prediction of the electron
$g$$-$$2$ of 2.4 ppb. This value is much larger than the 0.7~ppb relative
uncertainty very recently achieved in the measurement of
$a_e$~\cite{Gabrielse_g_2006}. The uncertainties in $A_2^{(4)}(m_e/m_{\mu})$
and $A_2^{(4)}(m_e/m_{\tau})$ are dominated by those in the latter and were
added in quadrature. The resulting error $\delta C^{(4)}_e = 6 \! \times \!
10^{-13}$ leads to a totally negligible $O(10^{-18})$ uncertainty in the
$a_e^{\mysmall \rm QED}$ prediction.

More than one hundred diagrams are involved in the evaluation of the
three-loop (sixth-order) {\small QED} contribution. Their analytic
computation required approximately three decades, ending in the late 1990s.
The coefficient $A_1^{(6)}$ arises from 72 diagrams. Its exact expression is
mainly due to Remiddi and his collaborators~\cite{Remiddi,LR96}; its
numerical value is
\be
     A_1^{(6)} = 1.181 \, 241 \, 456 \, 587 \ldots,
\label{eq:A16}
\ee
in very good agreement with previous results obtained with numerical
methods~\cite{Ki90-95}.

The calculation of the exact expression for the coefficient
$A_2^{(6)}(m_l/m_j)$ for arbitrary values of the mass ratio $m_l/m_j$ was
completed in 1993 by Laporta and Remiddi~\cite{La93,LR93} (earlier works
include Refs.~\cite{A26early}).  Let us focus on $a_e^{\mysmall \rm QED}$
($l\!=\!e$, $j\!=\!\mu$,$\tau$). This coefficient can be further split into
two parts: the first one, $A_2^{(6)}(m_l/m_j,\mbox{vac})$, receives
contributions from 36 diagrams containing either muon or $\tau$ vacuum
polarization loops~\cite{La93}, whereas the second one,
$A_2^{(6)}(m_l/m_j,\mbox{lbl})$, is due to 12 light-by-light scattering
diagrams with either muon or $\tau$ loops~\cite{LR93}. 
The exact expressions for these coefficients are rather complicated,
containing hundreds of polylogarithmic functions up to fifth degree (for the
light-by-light diagrams) and complex arguments (for the vacuum polarization
ones)---they also involve harmonic polylogarithms~\cite{HarmPol}. Series
expansions were provided in Ref.~\cite{LR93} for the cases of physical
relevance.

Using the recommended mass ratios given in Sec.~\ref{sec:QED}, the following
values were recently computed from the full analytic expressions~\cite{MP06}:
\bea
     A_2^{(6)}(m_e/m_{\mu}) &=& -7.373 \, 941 \, 64 \,(29) 
     \times 10^{-6}
\label{eq:EA26m}
     \\
     A_2^{(6)}(m_e/m_{\tau})&=& -6.5819 \,(19) 
     \times 10^{-8}.
\label{eq:EA26t}
\eea
Equations (\ref{eq:EA26m}) and (\ref{eq:EA26t}) provide the first evaluation
of the full analytic expressions for these coefficients with the {\small
CODATA} mass ratios of \cite{CODATA02}; they are almost identical to the
results $A_2^{(6)}(m_e/m_{\mu}) \!=\! -7.373 \, 941 \, 58 (28) \!\times\!
10^{-6}$ and $A_2^{(6)}(m_e/m_{\tau}) \!=\! -6.5819(19) \!\times\! 10^{-8}$
obtained in \cite{CODATA02} via the approximate series expansions in the
mass ratios. The small difference between $A_2^{(6)}(m_e/m_{\mu})$ of
\cite{CODATA02} and \eq{EA26m} mainly origins from the $O((m_e/m_{\mu})^6)$
term in the series expansion of $A_2^{(6)}(m_e/m_{\mu},\mbox{lbl})$; indeed,
due to its smallness, this term was neglected in the expansions of
Ref.~\cite{LR93} used in \cite{CODATA02}. The $O((m_e/m_{\mu})^6)$ and
$O((m_e/m_{\mu})^8)$ terms were explicitly provided in Ref.~\cite{MP06}
expanding the exact Laporta--Remiddi expression for the sum of
light-by-light and vacuum polarization contributions for $m_l/m_{j}
\!\ll\!  1$ (see also \cite{La93,KOPV} for parts of these expressions).  The
value of $A_2^{(6)}(m_e/m_{\mu})$ obtained including these additional terms
perfectly agrees with that in Eq.~(\ref{eq:EA26m}) determined with the exact
formulae. Indeed, their difference is of $O(10^{-23})$, to be compared with
the $O(10^{-13})$ error $\delta A_2^{(6)}(m_e/m_{\mu})$ due to the present
uncertainty of the ratio $m_e/m_{\mu}$.
Therefore, it will be possible to compute $A_2^{(6)}(m_e/m_{\mu})$ with the
simple expansion in Ref.~\cite{MP06}---thus avoiding the complexities of the
exact expressions---even if the precision of the ratio $m_e/m_{\mu}$ will
improve in the future by orders of magnitude.

The contribution of the three-loop diagrams with both $\mu$ and $\tau$ loop
insertions in the photon propagator can be calculated numerically from the
integral expressions of Ref.~\cite{Samuel91}. The value reported in
Ref.~\cite{MP06} is
\be
     A_3^{(6)}(m_e/m_{\mu},m_e/m_{\tau}) = 1.909 \, 45 \,(62) \times 10^{-13},
\label{eq:EA36}
\ee
a totally negligible $O(10^{-21})$ contribution to $a_{e}^{\mysmall \rm
QED}$.  Adding up Eqs.~(\ref{eq:A16}), (\ref{eq:EA26m}), (\ref{eq:EA26t})
and (\ref{eq:EA36}) one obtains the three-loop {\small QED}
coefficient~\cite{MP06}
\be
    C^{(6)}_e    =   1.181 \, 234 \, 016 \, 827 \,(19).
\label{eq:EC3}
\ee
The relative contribution to $a_{e}^{\mysmall \rm QED}$ of the
mass-dependent part of this three-loop coefficient is ${\sim} 0.1\,$ppb. This
is smaller than the present ${\sim} 0.7\,$ppb experimental
uncertainty~\cite{Gabrielse_g_2006}.  The error $1.9 \! \times \! 10^{-11}$
in \eq{EC3} leads to a totally negligible $O(10^{-19})$ uncertainty in
$a_e^{\mysmall \rm QED}$.

\subsection{Determination of \boldmath $\alpha$ from $a_e$ \unboldmath}
\label{subsec:ALPHA}

\noindent
As we already mentioned, recently a new measurement of the electron
anomalous magnetic moment by Gabrielse and his collaborators achieved the
extremely small relative uncertainty of $0.7\,$ppb~\cite{Gabrielse_g_2006},
\be 
a_{e}^{\mysmall \rm EXP} = 115 \, 965 \, 218\, 0.85 \, (76) \times 10^{-12}.
\label{eq:aEEX}
\ee
This uncertainty is nearly six times smaller than that of the last
measurement of $a_e$ reported back in 1987, $a_{e}^{\mysmall \rm EXP} =
1159652188.3 (4.2) \!\times\!  10^{-12}$~\cite{UW87,CODATA02}. These two
measurements differ by 1.7 standard deviations.

The fine-structure constant $\alpha$ can be determined equating the
theoretical {\small SM} prediction of the electron $g$$-$$2$ with its
measured value
\be 
      a_{e}^{\mysmall \rm SM}(\alpha) = a_{e}^{\mysmall \rm EXP}.
\label{eq:SMvsEXP}
\ee
The {\small SM} prediction contains the {\small QED} contribution
$a_{e}^{\mysmall \rm QED}(\alpha)=\sum_{i=1}^5 C^{(2i)}_e (\alpha/\pi)^i$ 
(higher-order coefficients are assumed to be negligible), plus small weak
and hadronic loop effects:
$a_{e}^{\mysmall \rm SM}(\alpha) = a_{e}^{\mysmall \rm QED}(\alpha) +
 a_{e}^{\mysmall \rm EW}  + a_{e}^{\mysmall \rm HAD}$ 
(the dependence on $\alpha$ of any contribution other than $a_{e}^{\mysmall
\rm QED}$ is negligible). The {\small EW} contribution is~\cite{CODATA02}:
\be
     a_{e}^{\mysmall \rm EW} = 0.0297 \, (5) \times 10^{-12};
\label{eq:aEEW}
\ee
this precise value includes the two-loop contributions calculated in
Refs.~\cite{CKM95L,CKM95D,CK96}. The hadronic term
is~\cite{CODATA02,AEHDR,Krause96}:
\be
     a_{e}^{\mysmall \rm HAD} = 1.671 \, (19)\times 10^{-12}.
\label{eq:aEHD}
\ee
The latest value for the four-loop {\small QED} coefficient is
$C^{(8)}_e \!\!=\!-1.7283 (35)$~\cite{KN2005-4LE}.
Following the argument of~\cite{CODATA02}, the educated guess
$C^{(10)}_e \!\!=\!0.0 (3.8)$ for the five-loop coefficient can be adopted.
The errors $\delta C^{(8)}_e\!=\!0.0035$ and $\delta C^{(10)}_e \!=\! 3.8$
lead to an uncertainty of $0.1 \!\times\! 10^{-12}$ and $0.3 \!\times\!
10^{-12}$ in $a_e^{\mysmall \rm QED}$, respectively. Solving \eq{SMvsEXP}
with the new measured value of \eq{aEEX}, one
obtains~\cite{Gabrielse_a_2006,MP06}
\bea
\alpha^{-1} \, &=& \, 137.035 \, 999 \, 709 \, (12)\,(30)\,(2)\,(90) 
               \nonumber \\
               &=& \, 137.035 \, 999 \, 709 \, (96)\,[0.70\,\mbox{ppb}]. 
\label{eq:alpha}
\eea
The first and second errors are due to the uncertainties of the four- and
five-loop {\small QED} coefficient $\delta C^{(8)}_e$ and $\delta
C^{(10)}_e$, respectively; the third one is caused by the tiny $\delta
a_{e}^{\mysmall \rm HAD}$, and the last one ($90 \!\times\! 10^{-9}$) is
from the experimental $\delta a_{e}^{\mysmall \rm EXP}$ in \eq{aEEX}.  The
uncertainties of the {\small EW} and two/three-loop {\small QED}
contributions are totally negligible at present.

The amazing precision of the determination in \eq{alpha} represents the
first significant improvement of this fundamental constant in a decade.  At
present, the best determinations of $\alpha$ independent of the electron
$g$$-$$2$ are
\bea
\alpha^{-1} ({\rm Rb}) &=& 137.035 \, 998 \, 78 \, (91)\,[6.7\,\mbox{ppb}], 
\label{eq:alphaRb}
\\
\alpha^{-1} ({\rm Cs}) &=& 137.036 \, 000 \, 0 \, (11) \,[8.0\,\mbox{ppb}];
\label{eq:alphaCs}
\eea
they are less precise by roughly a factor of ten.  The value $\alpha^{-1}
({\rm Rb})$ was deduced from the measurement of the ratio $h/M_{\rm Rb}$
based on Bloch oscillations of Rb atoms in an optical lattice ($h$ is the
Planck constant and $M_{\rm Rb}$ is the mass of the Rb atom)~\cite{Rb2006},
while $\alpha^{-1} ({\rm Cs})$ was determined from the measurement of the
ratio $h/M_{\rm Cs}$ ($M_{\rm Cs}$ is the mass of the Cs atom) via cesium
recoil measurement techniques~\cite{Wicht2002,Cs2006}. These two
determinations of $\alpha$ also rely on the precisely known Rydberg constant
and relative atomic masses of the electron, Rb and Cs
atoms~\cite{CODATA02,RelMassesRbCs}.  The values of $\alpha$ in
Eqs.~(\ref{eq:alphaRb}) and (\ref{eq:alphaCs}) are in good agreement with
the result of \eq{alpha}, differing from the latter by $-1.0$ and $+0.3$
standard deviations, respectively. This comparison provides a beautiful test
of the validity of {\small QED}. It also probes for possible electron
substructure~\cite{Gabrielse_a_2006}.

\section{MUON}
\label{sec:MUON}

In this section we will summarize the present status of the {\small SM}
prediction of the muon $g$$-$$2$. See
Refs.~\cite{Gminus2Reviews,DM04,MP04} for recent reviews.

\subsection{QED and Electroweak Contributions}

As we discussed in Secs.~\ref{sec:QED}--\ref{sec:ELECTRON}, the one-, two-,
and three-loop {\small QED} contributions to $a_{l}$ ($l\!=\!e$, $\mu$ or
$\tau$) are known analytically. The four-loop {\small QED} part of the muon
$g$$-$$2$, which is about six times larger than the present experimental
uncertainty of $\amu$, has thus far been evaluated only numerically. This
formidable task was first accomplished by Kinoshita and his collaborators in
the early 1980s~\cite{KL81-KNO84}.  The latest analyses appeared recently
in~\cite{KN2005-4LE,KN04-05}. The leading five-loop {\small QED} terms were
recently evaluated in~\cite{KN05-5loop} and were presented at this
workshop~\cite{NioTau06}; estimates obtained with the renormalization-group
method agree with these results~\cite{Kataev}. The total {\small QED}
prediction currently stands at
$ a_{\mu}^{\mysmall \rm QED} = 116 \, 584 \, 718.09 \, (14)\,(08) \times
     10^{-11}~\cite{MP06}.  
$
The first error is determined by the uncertainties of the {\small QED}
coefficients (dominated by the five-loop one), while the second is caused by
the tiny uncertainty of the recent value of the fine-structure constant
$\alpha$~\cite{Gabrielse_a_2006,MP06}.

The {\small EW} contribution to $\amu$ is suppressed by a
factor $(m_{\mu}/\mw)^2$ with respect to the {\small QED} effects. The
one-loop part was computed in 1972 by several authors~\cite{ew1loop}:
$
     \amu^{\mysmall \rm EW} (\mbox{1 loop}) = 
     \frac{5 G_{\mu} m^2_{\mu}}{24 \sqrt{2} \pi^2}
     \left[ 1+ \frac{1}{5}\left(1-4\sin^2\!\theta_{\mysmall{W}}\right)^2 
       + O(m^2_{\mu}/M^2_{\mysmall{Z,W,H}}) \right],
$
where $G_{\mu}=1.16637(1) \times 10^{-5}\gev^{-2}$ is the Fermi coupling
constant and $\theta_{\mysmall{W}}$ is the weak mixing angle.  Closed
analytic expressions for $\amu^{\mysmall \rm EW} (\mbox{1 loop})$ taking
exactly into account the $m^2_{\mu}/M^2_{\mysmall{B}}$ dependence ($B=Z,W,$
Higgs, or other hypothetical bosons) can be found in
Refs.~\cite{Studenikin}.  Employing the on-shell definition
$\sin^2\!\theta_{\mysmall{W}} =
1-M^2_{\mysmall{W}}/M^2_{\mysmall{Z}}$~\cite{Si80}, where
$\mz=91.1875(21)\gev$ and $\mw$ is the {\small SM} prediction of the $W$
mass (which can be derived, for example, from the simple formulae of
\cite{Formulette} leading to $\mw =80.383\gev$ for the Higgs mass
$\mh=150\gev$), one obtains $\amu^{\mysmall \rm EW} (\mbox{1 loop}) = 194.8
\times 10^{-11}$.

The two-loop {\small EW} contribution to $\amu$ is not negligible because of
large factors of $\ln(M_{\mysmall{Z,W}}/m_f)$, where $m_f$ is a fermion mass
scale much smaller than $\mw$~\cite{KKSS}. It was computed in
1995~\cite{CKM95L,CKM95D}.  The proper treatment of the contribution of the
light quarks was addressed in~\cite{PPD95KPPD02,CMV03}. These refinements
significantly improved the reliability of the fermionic part (that
containing closed fermion loops) of $\amu^{\mysmall \rm
EW}(\mbox{two-loop})$ leading, for $\mh=150\gev$, to
$
    \amu^{\mysmall \rm EW} = 154(1)(2)\times 10^{-11}
$~\cite{CMV03}.
The first error is due to hadronic loop uncertainties, while the second one
corresponds to an allowed range of $\mh \in [114,250]\gev$, to the current
top mass uncertainty, and to unknown three-loop effects.  The
leading-logarithm three-loop contribution to $\amu^{\mysmall \rm EW}$ is
extremely small~\cite{CMV03,DGi98}.
The result of \cite{HSW04} for the two-loop bosonic part of $\amu^{\mysmall
\rm EW}$, performed without the large $\mh$ approximation previously
employed in Ref.~\cite{CKM95L}, agrees with this previous evaluation in the
large Higgs mass limit.

\subsection{The Hadronic Contribution}

The evaluation of the hadronic leading-order contribution
$\amu^{\mbox{$\scriptscriptstyle{\rm HLO}$}}$, due to the hadronic vacuum
polarization correction to the one-loop diagram, involves long-distance
{\small QCD} for which perturbation theory cannot be employed. However,
using analyticity and unitarity, it was shown long ago that this term can be
computed from hadronic $e^+ e^-$ annihilation data via the dispersion
integral~\cite{DISP}
\be
      \amu^{\mbox{$\scriptscriptstyle{\rm HLO}$}}= 
      \frac{1}{4\pi^3} \!
      \int^{\infty}_{4m_{\pi}^2} ds \, K_{\mu}(s) \sigma^{(0)}\!(s),
\label{eq:dispint}
\ee
where $\sigma^{(0)}\!(s)$ is the total cross section for $e^+ e^-$
annihilation into any hadronic state, with extraneous {\small QED}
corrections subtracted off. The kernel function $K_{\mu}(s)$ decreases
monotonically for increasing~$s$.

A prominent role among all $e^+ e^-$ annihilation measurements is played by
the precise data collected in 1994-95 by the {\small CMD-2} detector at the
{\small VEPP-2M} collider in Novosibirsk for the $e^+e^-\rightarrow
\pi^+\pi^-$ cross section at values of $\sqrt{s}$ between 0.61 and 0.96
${\rm GeV}$~\cite{Akhmetshin:2003zn} (quoted systematic error 0.6\%,
dominated by the uncertainties in the radiative corrections).
Recently~\cite{CMD2-06} the {\small CMD-2} Collaboration released its
1996-98 measurements for the same cross section in the full energy range
$\sqrt{s} \in [0.37,1.39]$ GeV. The part of these data for $\sqrt{s} \in
[0.61,0.96]$ GeV (quoted systematic error 0.8\%) agrees with their earlier
result published in~\cite{Akhmetshin:2003zn}.
In 2005, also the {\small SND} Collaboration (at the {\small VEPP-2M}
collider as well) released its analysis of the $e^+e^-\rightarrow
\pi^+\pi^-$ process for $\sqrt{s}$ between 0.39 and 0.98 ${\rm GeV}$, with a
systematic uncertainty of 1.3\% (3.2\%) for $\sqrt{s}$ larger (smaller) than
0.42 GeV~\cite{Achasov:2005rg}. However, a recent reanalysis of these
data~\cite{NewSND} uncovered an error in the treatment of the radiative
corrections, reducing the value of the measured cross section.  The new
{\small SND} result appears to be in good agreement with the corresponding
one from {\small CMD2}.

In 2004 the {\small KLOE} experiment at the {\small DAFNE} collider in
Frascati presented a precise measurement of $\sigma(e^+e^-\rightarrow
\pi^+\pi^-)$~\cite{Aloisio:2004bu} via the initial-state radiation ({\small
ISR}) method~\cite{RadRet} at the $\phi$ resonance.  This cross section was
extracted for $\sqrt{s}$ between 0.59 and 0.97 GeV with a systematic error
of 1.3\% and a negligible statistical one.  There are some discrepancies
between the {\small KLOE} and {\small CMD-2} results, even if their
integrated contributions to $a_{\mu}^{\mbox{$\scriptscriptstyle{\rm HLO}$}}$
are similar. The data of {\small KLOE}~\cite{Aloisio:2004bu} and {\small
SND}~\cite{NewSND} particularly disagree above the $\rho$ peak, where the
latter are significantly higher.
The study of the $e^+e^-\rightarrow \pi^+\pi^-$ process via the {\small ISR}
method is also in progress at {\small BABAR}~\cite{BABAR-ISR} and
Belle~\cite{BELLE-ISR}.  On the theoretical side, analyticity, unitarity and
chiral symmetry provide strong constraints for the pion form factor in the
low-energy region~\cite{Colangelo}. Perhaps, also lattice {\small QCD}
computations of $\amu^{\mbox{$\scriptscriptstyle{\rm HLO}$}}$, although not
yet competitive with the precise results of the dispersive method, may
eventually rival that precision~\cite{LATTICE}.

Recent evaluations of the dispersive integral are in rather  good agreement:
\bea
      a_{\mu}^{\mbox{$\scriptscriptstyle{\rm HLO}$}} \!\!&=&\!\!\!
      6909 (39)_{exp} (19)_{rad} (7)_{qcd} \!\times\! 10^{-11}
\cite{Simon-Davier06}
\label{eq:DEHZ06}
\\
      a_{\mu}^{\mbox{$\scriptscriptstyle{\rm HLO}$}} \!\!&=&\!\!\!
      6894 \, (42)_{exp} (18)_{rad} \times 10^{-11}
~~\cite{Hagiwara:2006} 
\label{eq:HMNT06}
\\
      a_{\mu}^{\mbox{$\scriptscriptstyle{\rm HLO}$}} \!\!&=&\!\!\!
      6921 \, (56) \times 10^{-11} 
~~\cite{Jegerlehner} 
\label{eq:J06}
\\
      a_{\mu}^{\mbox{$\scriptscriptstyle{\rm HLO}$}} \!\!&=&\!\!\!
      6944 \, (48)_{exp} (10)_{rad} \times 10^{-11}
~~\cite{dTY04}.
\label{eq:TY04}
\eea
The three results in Eqs.~(\ref{eq:DEHZ06})--(\ref{eq:J06}) already include
the recent datasets from {\small CMD-2}~\cite{CMD2-06}, {\small
SND}~\cite{NewSND} and {\small BABAR}~\cite{babr} (not all of them, for the
value in \eq{J06}).
{\small KLOE}'s data were not included in the analyses of
Refs.~\cite{Simon-Davier06} and \cite{dTY04}.
The preliminary value in \eq{DEHZ06} updates the one of Ref.~\cite{DEHZ03}.
Significant progress is expected from the $e^+e^-$ collider {\small
VEPP-2000} under construction in Novosibirsk~\cite{VEPP2000} and, possibly,
from {\small DAFNE-2} at Frascati~\cite{DAFNE2}.

The authors of \cite{ADH98} pioneered the idea of using vector spectral
functions derived from the study of hadronic $\tau$
decays~\cite{DHZ05,PortolesTAU06} to improve the evaluation of the
dispersive integral.  However, the latest analysis with {\small ALEPH},
{\small CLEO}, and {\small OPAL} data yields
$a_{\mu}^{\mbox{$\scriptscriptstyle{\rm HLO}$}} = 7110 \, (50)_{exp}
(8)_{rad} (28)_{SU(2)} \times 10^{-11}$~\cite{DEHZ03},
a value significantly higher than those obtained with $e^+e^-$ data.
Isospin-breaking corrections were applied~\cite{MS88CEN}. 
Although the precise {\small CMD-2} $e^+e^-\rightarrow \pi^+\pi^-$ data are
consistent with the corresponding $\tau$ ones for energies below ${\sim}0.85$
GeV, they are significantly lower for larger energies. {\small KLOE}'s
$\pi^+\pi^-$ spectral function confirms this discrepancy with the $\tau$
data. Even if {\small SND}'s 2005 results~\cite{Achasov:2005rg} were
compatible with the $\tau$ ones, their recent reanalysis~\cite{NewSND},
reducing the value of the measured cross section, seems to indicate that
this is no longer the case.  Interesting preliminary results from Belle of
the study of the decay $\tau^- \to \pi^- \pi^0 \nu_{\tau}$ were presented at
this workshop~\cite{BELLE-TAU,FujikawaTAU06}.
This puzzling discrepancy between the $\pi^+\pi^-$ spectral functions from
$e^+e^-$ and isospin-breaking-corrected $\tau$ data, also discussed at this
workshop~\cite{CVC06}, could be caused by inconsistencies in the $e^+e^-$ or
$\tau$ data, or in the isospin-breaking corrections which must be applied to
the latter~\cite{eetau}.

The hadronic higher-order $(\alpha^3)$ contribution
$a_{\mu}^{\mbox{$\scriptscriptstyle{\rm HHO}$}}$ can be divided into two
parts:
$
     a_{\mu}^{\mbox{$\scriptscriptstyle{\rm HHO}$}}=
     a_{\mu}^{\mbox{$\scriptscriptstyle{\rm HHO}$}}(\mbox{vp})+
     a_{\mu}^{\mbox{$\scriptscriptstyle{\rm HHO}$}}(\mbox{lbl}).
$
The first one is the $O(\alpha^3)$ contribution of diagrams containing
hadronic vacuum polarization insertions~\cite{Krause96}. Its latest
value is $a_{\mu}^{\mbox{$\scriptscriptstyle{\rm HHO}$}}(\mbox{vp})=
-97.9 \, (0.9)_{exp} (0.3)_{rad} \times 10^{-11}
$~\cite{Hagiwara:2003da,Hagiwara:2006}; it changes by ${\sim} -3\times
10^{-11}$ if hadronic $\tau$-decay data are used instead~\cite{DM04}.
The second term, also of $O(\alpha^3)$, is the hadronic light-by-light
contribution. As it cannot be directly determined via a dispersion
relation approach using data (unlike the hadronic vacuum polarization
contribution), its evaluation relies on specific models of low-energy
hadronic interactions with electromagnetic currents. Three major
components of $a_{\mu}^{\mbox{$\scriptscriptstyle{\rm
HHO}$}}(\mbox{lbl})$ can be identified: charged-pion loops, quark
loops, and pseudoscalar ($\pi^0$, $\eta$, and $\eta'$) pole
diagrams~\cite{lbl1}.  The latter ones dominate the final result and
require information on the electromagnetic form factors of the
pseudoscalars. In 2001 the authors of~\cite{lbl2} uncovered a sign
error in earlier evaluations of the dominating pion-pole part. Their
estimate of $a_{\mu}^{\mbox{$\scriptscriptstyle{\rm
HHO}$}}(\mbox{lbl})$, based also on the previous results for the quark and
charged-pions loop parts, is
$
      a_{\mu}^{\mbox{$\scriptscriptstyle{\rm HHO}$}}(\mbox{lbl}) =  
      80\,(40)\times 10^{-11}
$.
A higher value was obtained in 2003 including short-distance
{\small QCD} constraints:
$
      a_{\mu}^{\mbox{$\scriptscriptstyle{\rm HHO}$}}(\mbox{lbl}) =
      136\,(25)\times 10^{-11}
$~\cite{MV03}.
The upper bounds found in Refs.~\cite{Erler:2006vu} are consistent
with this higher value.  Further independent calculations would
provide an important check of these results for
$\amu^{\mbox{$\scriptscriptstyle{\rm HHO}$}}(\mbox{lbl})$ (see
\cite{BP07} for a recent critical comparison of these evaluations), a
contribution whose uncertainty may become the ultimate limitation of
the {\small SM} prediction of the muon $g$$-$$2$.

\subsection{Standard Model vs.\ Measurement}

The first column of Table~\ref{table:n1} shows
$
    \amu^{\mysmall \rm SM} = 
         \amu^{\mysmall \rm QED} +
         \amu^{\mysmall \rm EW}  +
         \amu^{\mbox{$\scriptscriptstyle{\rm HLO}$}} +
         \amu^{\mbox{$\scriptscriptstyle{\rm HHO}$}}.
$
The values employed for $\amu^{\mbox{$\scriptscriptstyle{\rm HLO}$}}$ are
indicated by the reference on the left; the value of Ref.~\cite{DEHZ03}
was obtained using data from hadronic $\tau$ decays. All $\amu^{\mysmall \rm
SM}$ values were derived with $\amu^{\mbox{$\scriptscriptstyle{\rm
HHO}$}}(\mbox{lbl})\!=\!  80\,(40)\times 10^{-11}$~\cite{lbl1,lbl2}. Errors
were added in quadrature. 
\begin{table}
\caption{$a_{\mu}$: Standard Model vs.\ measurement.\protect\vphantom{fg}}
\label{table:n1}
\begin{center}
\begin{tabular}{lll}
\toprule
$\amu^{\mbox{$\scriptscriptstyle{\rm SM}$}} \times 10^{11}$ & 
$\Delta \times 10^{11}$  & ~~$\sigma$           \\
\midrule
\mbox{\cite{Simon-Davier06}}~~116\,591\,763 (60)~& 
                      317 (87) & ~~3.7 $\langle 3.2 \rangle$  \\
\mbox{\cite{Hagiwara:2006}}~~116\,591\,748 (61)~& 
                      332 (88) & ~~3.8 $\langle 3.4 \rangle$  \\
\mbox{\cite{Jegerlehner}}~~116\,591\,775 (69)~& 
                      305 (93)& ~~3.3 $\langle 2.8 \rangle$  \\
\mbox{\cite{dTY04}}~~116\,591\,798  (63)~& 
                      282 (89) & ~~3.2 $\langle 2.7 \rangle$  \\
\midrule
\mbox{\cite{DEHZ03}}~~116\,591\,961 (70)~&
                      119 (95) & ~~1.3 $\langle 0.7 \rangle$\\
\bottomrule
\end{tabular}
\end{center}
\end{table} 
The present world average experimental value is
$
    a_{\mu}^{\mysmall \rm EXP}  =
               116 \, 592 \, 080 \, (63) \times 10^{-11}
               ~[0.5~\mbox{ppm}]
$~\cite{BNL04-6}.
Note that the theoretical error on $a_{\mu}$ is now roughly the same as the
experimental one.  The differences $\Delta
=\amu^{\mbox{$\scriptscriptstyle{\rm EXP}$}}-
\amu^{\mbox{$\scriptscriptstyle{\rm SM}$}}$ are listed in the second column
of Table~\ref{table:n1}, while the numbers of ``standard deviations''
($\sigma$) appear in the third one. Lower discrepancies, shown in angle
brackets, are obtained if $\amu^{\mbox{$\scriptscriptstyle{\rm
HHO}$}}(\mbox{lbl}) = 136\,(25)\times 10^{-11}$~\cite{MV03} is used instead
of $80\,(40)\times 10^{-11}$~\cite{lbl1,lbl2}.

\section{TAU}
\label{sec:TAU}

This section summarizes the results reported at this workshop~\cite{EGIP06}
for the theoretical prediction of the $\tau$ lepton $g$$-$$2$.  We also
refer the reader to Ref.~\cite{EP06} for a very recent update and review.

\subsection{QED and Electroweak Contributions}

The sum of the one-, two- and three-loop {\small QED} contribution to
$a_{\tau}$, obtained using the mass ratios given in Sec.~\ref{sec:QED}
is~\cite{MP06}:
\be
    a_{\tau}^{\mysmall \rm QED} =
    117 \, 324 \, (2) \times 10^{-8}.
\label{eq:TQED}
\ee
The error $\delta a_{\tau}^{\mysmall \rm QED}$ is the uncertainty
$\delta C_{\tau}^{(8)}(\alpha/\pi)^4 \sim \pi^2 \ln^2(m_{\tau}/m_e)
(\alpha/\pi)^4 \sim 2\times 10^{-8}$
assigned in \cite{MP06} to $a_{\tau}^{\mysmall \rm QED}$ for uncalculated
four-loop contributions. The errors due to the uncertainties of the
$O(\alpha^2)$ ($5 \times 10^{-10}$) and $O(\alpha^3)$ terms ($3 \times
10^{-11}$), as well as that induced by the uncertainty of $\alpha$ ($8
\times 10^{-13}$) are negligible.
The result in \eq{TQED} supersedes the earlier values $a_{\tau}^{\mysmall
\rm QED} = 117 \, 319 \, (1) \times 10^{-8}$~\cite{Samuel_tau} and
$a_{\tau}^{\mysmall \rm QED} = 117 \, 327.1 \, (1.2) \times
10^{-8}$~\cite{Narison01} (see Refs.~\cite{MP06,EP06} for details).

The {\small EW} correction to the anomalous magnetic moment of the $\tau$
lepton is of the same order of magnitude as the three-loop {\small QED}
one. The one-loop {\small EW} term is
$ a_{\tau}^{\mysmall \rm EW} (\mbox{1 loop}) = 55.1 \times
    10^{-8}~\cite{ew1loop}.
\label{eq:EWoneloopN}
$
The estimate of the total {\small EW} contribution of
Ref.~\cite{Samuel_tau}, $a_{\tau}^{\mysmall \rm EW} = 55.60(2) \times
10^{-8}$, obtained from the one-loop formula, is similar to the one-loop
value reported above. However, it doesn't contain the two-loop contribution,
which is not negligible.

As we already discussed for the muon $g$$-$$2$, the two-loop {\small EW}
contributions $a_l^{\mysmall \rm EW} (\mbox{2 loop})$ ($l\!=\!e$, $\mu$ or
$\tau$) were computed in 1995~\cite{CKM95L,CKM95D,CK96}, leading to a
significant reduction of the one-loop prediction because of large factors of
$\ln(M_{\mysmall{Z,W}}/m_f)$, where $m_f$ is a fermion mass scale much
smaller than $\mw$~\cite{KKSS}.  The numerical value of the bosonic part of
$a_{\tau}^{\mysmall \rm EW} (\mbox{2 loop})$, for $\mh=150\gev$, is
$a_{\tau}^{\mysmall \rm EW}(\mbox{2 loop bos}) = -3.06 \times
10^{-8}$~\cite{CKM95L,CK96,EGIP06,EP06}.
Very recently, the analysis of the fermionic part of Ref.~\cite{CKM95D,CK96}
was slightly refined in Ref.~\cite{EP06} leading, for $\mh\!=\!150\gev$, to
$a_{\tau}^{\mysmall \rm EW}(\mbox{2 loop fer}) = -4.68 \times 10^{-8}$.
The sum of the fermionic and bosonic two-loop {\small EW} contributions
gives
$a_{\tau}^{\mysmall \rm EW}(\mbox{2 loop}) \!=\! -7.74 \times
10^{-8}$~\cite{EGIP06,EP06},
a 14\% reduction of the one-loop result. As discussed in
Sec.~\ref{sec:MUON}, the leading-logarithm three-loop {\small EW}
contributions to the muon $g$$-$$2$ were determined to be extremely small
via renormalization-group analyses~\cite{CMV03,DGi98}; in
Ref.~\cite{EGIP06,EP06} an additional uncertainty of
$O[a_{\tau}^{\mysmall \rm EW}(\mbox{2 loop}) (\alpha/\pi)
\ln(\mz^2/m_{\tau}^2)] \!\sim\! O(10^{-9})$
was assigned to $a_{\tau}^{\mysmall \rm EW}$ to account for these neglected
three-loop effects. Adding $a_{\tau}^{\mysmall \rm EW}(\mbox{2 loop})$ to
the one-loop value presented above, one gets the total {\small EW}
correction ($\mh\!=\!150\gev$)~\cite{EGIP06,EP06}:
\be
    a_{\tau}^{\mysmall \rm EW} = 47.4 (5) \times 10^{-8}.
\label{eq:TEW}
\ee
The uncertainty allows $\mh$ to range from 114 GeV up to ${\sim} 300$ GeV,
and reflects the estimated errors induced by hadronic loop effects,
neglected two-loop bosonic terms and the missing three-loop contribution. It
also includes the tiny errors due to the uncertainties in
$M_{\rm\scriptstyle top}$ and $m_{\tau}$. The value in \eq{TEW} is in
agreement with the prediction
  $a_{\tau}^{\mysmall \rm EW} = 47 (1) \times 10^{-8}$~\cite{CK96,Narison01},
with a reduced uncertainty.  As we mentioned earlier, the {\small EW}
estimate of Ref.~\cite{Samuel_tau}, $a_{\tau}^{\mysmall \rm EW} = 55.60(2)
\times 10^{-8}$, mainly differs from \eq{TEW} in that it doesn't include the
two-loop corrections.

\subsection{The Hadronic Contribution}

Similarly to the case of the muon, the leading-order hadronic contribution
to the $\tau$ lepton $g$$-$$2$ is given by the dispersion integral in
\eq{dispint} (with the kernel $K_{\mu}(s)$ replaced by $K_{\tau}(s)$), in
which the role of the low energies is very important, although not as
strongly as in $a_{\mu}^{\mysmall \rm HLO}$.  The history of the
$a_{\tau}^{\mysmall \rm HLO}$
calculations~\cite{EP06,Samuel_tau,Narison01,nar78,bs88,ej95,j96} based
mainly on experimental $e^+e^-$ data is shown in Table~\ref{table:n2}.  Purely
theoretical estimates somewhat undervalue the hadronic contribution and have
rather large uncertainties~\cite{ben93,hn93,hold,dor}.
%

\vspace{-5mm}
\begin{table}[htb]
\caption{Calculations of ${\rm a}_{\tau}^{\mysmall \rm HLO}$.}
\label{table:n2}
\begin{center}
\begin{tabular}{@{}lc@{}} 
\toprule
Author & ${\rm a}_{\tau}^{\mysmall \rm HLO} \times 10^{8}$  \\
\midrule
Narison \cite{nar78} & 370 $\pm$ 40 \\
Barish \& Stroynowski~\cite{bs88} & ${\sim} 350$ \\
Samuel et al. \cite{Samuel_tau}  &  $360 \pm 32$ \\
Eidelman \& Jegerlehner \cite{ej95,j96}  & 
$338.4 \pm 2.0 \pm 9.1$ \\
Narison \cite{Narison01} &  353.6 $\pm$ 4.0 \\
Eidelman \& Passera~\cite{EGIP06,EP06}   & 337.5 $\pm$ 3.7 \\
\bottomrule
\end{tabular}
\end{center}
\end{table}
\vspace{-5mm}
The calculation of the leading-order contribution was very recently updated
using the whole bulk of experimental data below 12~GeV, which include, as
discussed in Sec.~\ref{sec:MUON}, old data compiled in
Refs.~\cite{ej95,DEHZ03}, and recent results from {\small
CMD-2}~\cite{CMD2-06}, {\small SND}~\cite{NewSND}, {\small
KLOE}~\cite{Aloisio:2004bu} and {\small BABAR}~\cite{babr}. The improvement
is particularly strong in the channel $e^+e^- \to\pi^+\pi^-$.  The result of
this analysis is~\cite{EGIP06,EP06}:
\be
    a_{\tau}^{\mysmall \rm HLO} =  337.5 \, (3.7) \times 10^{-8}
\label{eq:THLO}.
\ee 
The overall uncertainty is 2.5 times smaller than that of the previous
data-based prediction~\cite{ej95,j96}.

Like $a_{\mu}^{\mysmall \rm HHO}$, the hadronic higher-order $(\alpha^3)$
contribution $a_{\tau}^{\mysmall \rm HHO}$ can be divided into two parts:
$
     a_{\tau}^{\mysmall \rm HHO}=
     a_{\tau}^{\mysmall \rm HHO}(\mbox{vp})+
     a_{\tau}^{\mysmall \rm HHO}(\mbox{lbl}).
$
The first one, the $O(\alpha^3)$ contribution of diagrams containing
hadronic self-energy insertions in the photon propagators,
is~\cite{Krause96}:
\be
a_{\tau}^{\mysmall \rm HHO}(\mbox{vp})= 7.6 (2) \times 10^{-8}.
\label{eq:THHOVAC}
\ee
Note that na\"{\i}vely rescaling the muon result by the factor
$m_{\tau}^2/m_{\mu}^2$ leads to the totally incorrect estimate
$a_{\tau}^{\mysmall \rm HHO}(\mbox{vp})= (-101\times 10^{-11}) \times
m_{\tau}^2/m_{\mu}^2 = -29 \times 10^{-8}$ (the $a_{\mu}^{\mysmall \rm
HHO}(\mbox{vp})$ value is from Ref.~\cite{Krause96}); even the sign is
wrong! Until recently, very few estimates of the light-by-light contribution
$a_{\tau}^{\mbox{$\scriptscriptstyle{\rm HHO}$}}(\mbox{lbl})$ existed in the
literature~\cite{Samuel_tau,Narison01,Krause96}, and all of them were
obtained simply rescaling the muon results $a_{\mu}^{\mysmall \rm
HHO}(\mbox{lbl})$ by a factor $m_{\tau}^2/m_{\mu}^2$. These very na\"{\i}ve
estimates fall short of what is needed, as this scaling is not justified.
For these reasons, a parton-level estimate of $a_{\tau}^{\mysmall \rm
HHO}(\mbox{lbl})$ was recently performed in Ref.~\cite{EP06},
obtaining
\be
a_{\tau}^{\mysmall \rm HHO}(\mbox{lbl})= 5 (3) \times 10^{-8}.
\label{eq:THHOLBL}
\ee
This value is much lower than those obtained by simple rescaling. 
We refer the reader to Ref.~\cite{EP06} for a detailed discussion.

\subsection{Standard Model prediction for \bm $a_{\tau}$ \ubm}

We can now add up all the contributions discussed in the previous sections
to derive the {\small SM} prediction for $a_{\tau}$~\cite{EGIP06,EP06}:
\bea
    a_{\tau}^{\mysmall \rm SM} &=& 
         a_{\tau}^{\mysmall \rm QED} +
         a_{\tau}^{\mysmall \rm EW}  +
         a_{\tau}^{\mysmall \rm HLO}  +
         a_{\tau}^{\mysmall \rm HHO} \nonumber \\
         &=&117 \, 721 \, (5) \times 10^{-8}.  
\label{eq:nsm}
\eea
Errors were added in quadrature.
%
%
Comparing the most stringent experimental limit mentioned above, $ -0.052 <
a_{\tau}^{\mysmall \rm EXP} < 0.013$ at 95\% confidence level~\cite{delphi},
with \eq{nsm}, it is clear that the sensitivity of the best existing
measurements is still more than an order of magnitude worse than needed.
For other limits on $a_{\tau}$ see Ref.~\cite{PDG06,arcadi}.

\section{CONCLUSIONS AND OUTLOOK}

The study of the anomalous magnetic moment of the electron and the muon
provides a beautiful example of interplay between theory and experiment. It
started more than 50 years ago: Schwinger's 1948 calculation~\cite{Sch48} of
the leading contribution to $a_e$ was
one of the very first results of {\small QED}, and its agreement with the
experimental value provided one of the early confirmations of this
theory. In July 2006 Gabrielse and his collaborators published their
measurement of $a_e$ with an unprecedented accuracy of
$0.7\,$ppb~\cite{Gabrielse_g_2006}, breaking the 1987 world record held by
Dehmelt and his collaborators~\cite{UW87}. On the theoretical side, heroic
higher-order {\small QED} calculations were performed over the years by
Kinoshita, Remiddi, and their collaborators. A result of this extraordinary
synergy is today's determination of the fine-structure constant via $a_e$,
$\alpha^{-1} = 137.035 \, 999 \, 709 \, (96)$~\cite{Gabrielse_a_2006,MP06},
whose uncertainty is roughly ten times smaller than that from any
other method. A further reduction of the experimental error $\delta
a_e^{\mysmall \rm EXP}$, which dominates the uncertainty of $\alpha$,
appears to be at hand~\cite{Gabrielse_g_2006,Gabrielse_a_2006}.

The {\small SM} prediction of the muon $g$$-$$2$ deviates from the present
experimental value by more than $3\,\sigma$, if data from $e^+e^-$ collisions
are employed to evaluate the leading-order hadronic term (the recent
datasets from the {\small CMD-2}~\cite{CMD2-06}, {\small SND}~\cite{NewSND}
and {\small BABAR}~\cite{babr} are already included). This deviation cannot
be ignored. Determinations based on $\tau$-decay data deviate only by
approximately $1\,\sigma$.  The puzzling discrepancy between the $\pi^+\pi^-$
spectral functions from $e^+e^-$ and isospin-breaking-corrected $\tau$ data
could be caused by inconsistencies in the $e^+e^-$ or $\tau$ data, or in the
isospin-breaking corrections applied to the latter. Indeed, some
disagreements occur between $e^+e^-$ data sets, requiring further detailed
investigations. On the other hand, the connection of $\tau$ data with the
leading hadronic contribution to $\amu$ is less direct, and one wonders
whether all possible isospin-breaking effects have been properly taken into
account.

The impressive results of the {\small E821} experiment are still limited by
statistical errors.  A new experiment, {\small E969}, has been approved (but
not yet funded) at Brookhaven in 2004~\cite{E969}. Its goal is to reduce the
present experimental uncertainty by a factor of 2.5 to about 0.2 ppm. A
letter of intent for an even more precise muon $g$$-$$2$ experiment was
submitted to {\small J-PARC} with the proposal to reach a precision below
0.1 ppm~\cite{JPARC}. While the {\small QED} and {\small EW} contributions
appear to be ready to rival these precisions, much effort will be needed to
reduce the hadronic uncertainty by a factor of two. This effort is
challenging but possible, and certainly well motivated by the excellent
opportunity the muon $g$$-$$2$ is providing us to unveil (or constrain)
{\small NP} effects.

The $g$$-$$2$ of the $\tau$ lepton is even more sensitive than the muon one
to {\small EW} and {\small NP} loop effects that give contributions ${\sim}
m_l^2$. However, unfortunately, the very short lifetime of the $\tau$ lepton
makes it very difficult to determine $a_{\tau}$ by measuring its spin
precession in the magnetic field, like in the muon $g$$-$$2$
experiment~\cite{BNL,BNL04-6}. Instead, experiments focus on high-precision
measurements of the $\tau$ lepton pair production in various high-energy
processes and comparison of the measured cross sections with the {\small
QED} predictions~\cite{delphi,PDG06}, but their sensitivity is still more
than an order of magnitude worse than that required to determine $a_{\tau}$.

Nonetheless, there are many interesting suggestions to measure
$a_{\tau}$, e.g., from the radiation amplitude zero in radiative
$\tau$ decays~\cite{rad_zero} or from other observables; these methods
could possibly exploit the very large $\tau$ lepton samples collected
at $B$ factories. Reference \cite{samu} suggests a similar method to
study $a_{\tau}$ using radiative $W$ decays and potentially very high
data samples at {\small LHC}.
Yet another method would use the channeling
in a bent crystal similarly to the measurement of magnetic moments of
short-living baryons~\cite{bent_crystal}.  In the case of the $\tau$ lepton,
it was suggested to use the decay $B^+ \to \tau^+ \nu_{\tau}$, which would
produce polarized $\tau$ leptons~\cite{Samuel_tau} and was recently
observed~\cite{btau}.  We believe that a detailed feasibility study of such
experiments, as well as further attempts to improve the accuracy of the
theoretical prediction for $a_{\tau}$, are quite timely.


\section*{ACKNOWLEDGMENTS}

\noindent I would like to thank the organizers for their excellent
coordination of this workshop, and in particular A.\ Lusiani and S.\
Eidelman for their kind invitation. I am also grateful to S.\ Eidelman, M.\
Giacomini and F.V.\ Ignatov for many fruitful discussions and collaborations.



\end{document}